\documentclass[12pt,letterpaper]{article}
\pdfoutput=1
\usepackage{jheppub}
\usepackage{amsfonts,amsthm}
\usepackage[english]{babel}
\usepackage[utf8]{inputenc}
\usepackage{slashed}
\usepackage{mathrsfs}
\usepackage{amssymb}
\usepackage{color}
\hypersetup{unicode}
\usepackage{natbib}
\usepackage{comment}

\usepackage{tikz}
\usetikzlibrary{decorations.pathmorphing,arrows.meta}

\newcommand{\eq}{\begin{equation}}
\newcommand{\feq}{\end{equation}}
\newcommand{\be}{\begin{equation}}
\newcommand{\ee}{\end{equation}}
\newcommand{\eqn}{\begin{eqnarray}}
\newcommand{\feqn}{\end{eqnarray}}
\DeclareMathOperator{\sech}{sech}

\def\arcsinh{{\mathrm{arcsinh}}}

\title{R7-branes and $\text{AdS}_{\text{9}}$ solutions in type IIB}

\author[a]{Giuseppe Dibitetto}
\author[b]{Cameron Gibson}
\author[c]{Nicol\`o Petri}
\author[d]{Colin Sterckx}

\affiliation[a]{Dipartimento di Fisica, Università di Roma “Tor Vergata” \& INFN, Sezione di Roma 2, Via della
Ricerca Scientifica 1, 00133 Roma, Italy.}
\affiliation[b]{George P. \& Cynthia Woods Mitchell Institute for Fundamental Physics and Astronomy,
Texas A{\&}M University, College Station, TX 77843, USA.}
\affiliation[c]{INFN, Sezione di Milano, Via Celoria 16, 20133 Milano, Italy.}
\affiliation[d]{Universit\'e Libre de Bruxelles (ULB) and International Solvay Institutes,
Service de Physique Th\'eorique et Math\'ematique,
Campus de la Plaine, CP 231, B-1050, Brussels, Belgium.}

\emailAdd{giuseppe.dibitetto@roma2.infn.it}
\emailAdd{camerongibson@tamu.edu}
\emailAdd{nicolo.petri@mi.infn.it}
\emailAdd{colin.sterckx@ulb.ac.be}

\abstract{We consider the recently found $\mathrm{AdS}_9$ backgrounds in type IIB supergravity. We start by reviewing the solutions in a more convenient set of coordinates, which allows us to explicitly compute a few relevant observables, including the holographic central charge. Subsequently, we study non-supersymmetric 7-brane configurations sourced by the IIB axio-dilaton. We find explicit analytic 7-brane solutions that possess the above $\mathrm{AdS}_9$ backgrounds as their near-horizon geometry. Finally, we discuss the relation of these 7-brane solutions to the R7 branes by imposing the reflection monodromy $\tau \rightarrow -\bar \tau$ in the transverse plane.}

\begin{document}
`\maketitle
\flushbottom

\section{Introduction}

Anti-de Sitter backgrounds are among the most extensively studied solutions of string theory. In the best understood examples, they arise as the near-horizon geometry of branes. This relation provides a particularly clear realization of the AdS/CFT correspondence, as it directly connects the gravitational description to the microscopic degrees of freedom living on the brane worldvolume. D3-, M2-, and M5-branes provide the canonical examples.

This naturally leads to a broader question: can every AdS solution be understood as the near-horizon geometry of a brane? While this is well understood for the simplest and most symmetric AdS backgrounds, much less is known for warped AdS solutions (see for instance \cite{Cvetic:2000cj}), especially in the absence of supersymmetry. In the latter case, the microscopic origin of the corresponding supergravity solutions is not yet understood.

This question has acquired additional significance in light of recent developments in the Swampland program. The non-SUSY AdS conjecture proposes that stable non-supersymmetric AdS vacua cannot exist in a consistent theory of quantum gravity. This naturally raises the question of their microscopic realization and eventual decay \cite{Arkani-Hamed:2006emk,Ooguri:2016pdq}. At the same time, the Cobordism conjecture \cite{McNamara:2019rup} has revealed a much richer structure of non-BPS extended objects mediating transitions between different vacua and string theories \cite{Heckman:2025wqd,Torres:2026vxx,Anastasi:2026cus}. Taken together, these developments indicate that our understanding of non-supersymmetric vacua may require entirely new classes of non-perturbative configurations beyond those explored so far.

One particularly interesting example is the proposal of a new class of codimension-two defects in type IIB string theory, known as R7-branes \cite{Dierigl:2022reg,Debray:2023yrs,Dierigl:2023jdp,Heckman:2025wqd,Hamada}. These objects are characterized by a reflection monodromy of the axio-dilaton, which completely breaks supersymmetry. A first question is whether these defects admit a gravitational description. The first evidence in this direction was provided in \cite{Cavusoglu:2026xiv}, where perturbative 7-brane solutions with reflection monodromy were constructed in axio-dilaton gravity. In Appendix D of \cite{Cavusoglu:2026xiv}, the authors also observed that the AdS$_9$ geometry arises as a particular limit of their general solutions, providing the first indication of a possible connection between R7-branes and AdS$_9$.

Motivated by these developments, an analytic family of AdS$_9$ solutions in type IIB warped over a line and supported by an axio-dilaton was recently constructed in \cite{Dibitetto:2026oor}. These solutions belong to a much broader class of warped AdS$_D\times I$ backgrounds in axio-dilaton gravity in arbitrary dimensions \cite{Dibitetto:2026yft}. Even if they are analytic, the AdS$_9$ solutions have so far lacked a microscopic interpretation. A first step towards uncovering their microscopic origin is to determine whether they arise as the near-horizon geometry of a brane configuration, as originally suggested in \cite{Cavusoglu:2026xiv,Dibitetto:2026oor}.

The main purpose of this paper is to establish the connection between AdS$_9$ and R7-branes through an explicit construction of the corresponding 7-brane solutions. We emphasize that constructing local solutions describing non-supersymmetric branes is a highly non-trivial problem. Significant progress has recently been made in the context of non-supersymmetric string theories \cite{Dudas:2000ff,Mourad:2016xbk,Basile:2018irz,Sagnotti:2021mxb,Baykara:2022cwj,Raucci:2022bjw,Raucci:2022jgw,Mourad:2024dur,Raucci:2025bev,Basile:2026lyc}, as well as in de Sitter-like solutions \cite{Horowitz:2007pr,Ooguri:2017njy,GarciaEtxebarria:2020xsr,Bomans:2021ara,Giri:2021eob,Dibitetto:2022rzy,Menet:2025nbf,Andriot:2026lac,Ghodsi:2026lmn} and anti de Sitter solutions \cite{Guarino:2020flh,Giambrone:2021wsm}, and in the framework of Dynamical Cobordism \cite{Angius:2022aeq,Angius:2023uqk,Huertas:2023syg}. The AdS$_9$ solutions constructed in \cite{Dibitetto:2026oor} share several features with those of \cite{Cordova:2018eba,Cordova:2018dbb}, but, in contrast to that case, they admit an analytic description. Using this feature, in this paper we introduce a new coordinate system in which the AdS$_9$ solutions of \cite{Dibitetto:2026oor} take a particularly simple form. This reformulation allows us to construct a new family of exact 7-brane solutions of type IIB supergravity whose near-horizon geometry reproduces the AdS$_9$ backgrounds of \cite{Dibitetto:2026oor}. We then show that this solution admits an asymptotically Mink$_{10}$ geometry and the appropriate reflection monodromy. We thus propose to interpret this solution as the gravitational description of R7-branes.

The new parametrization also makes it possible to derive an exact analytic expression for the holographic central charge following \cite{Klebanov:2007ws,Cremonesi:2015bld}. We find that it scales as $(L/\ell_s)^8$ where $L$ and $\ell_s$ are the AdS and string scales. However, at present no dual field theory is known that could account for it. Explaining the origin of this scaling from a microscopic perspective therefore remains an important open problem.

Finally, we show that our 7-brane solutions admit a natural twelve-dimensional geometric uplift as Ricci-flat manifolds. Unlike the familiar supersymmetric case, however, these monopoles are intrinsically non-holomorphic, making their geometric interpretation considerably more subtle.

The paper is organized as follows. In Section \ref{section2}, we revisit the AdS$_9$ solutions of type IIB supergravity, discuss flux quantization, compute the holographic central charge, and analyze their relation to R7-branes. In Section \ref{section3}, we construct explicit analytic 7-brane solutions with an AdS$_9$ limit, asymptotically Mink$_{10}$ geometry and the appropriate R7-brane reflection monodromy. In Section \ref{addedsection}, we construct 7-brane solutions with the same properties, with the additional feature that AdS$_9$ limit corresponds to the Poincar\'{e} horizon instead of the conformal boundary. In Section \ref{section4}, we present the corresponding twelve-dimensional geometric description and discuss its interpretation. We conclude in Section \ref{conclusions} with a discussion of our results and future directions.

\section{A second look at $\text{AdS}_{\text{9}}$ solutions in type IIB supergravity}\label{section2}
Let us start by reviewing the $\mathrm{AdS}_9$ solutions of type IIB supergravity, which were recently found in \cite{Dibitetto:2026oor}. These backgrounds are obtained by simply assuming an $\mathrm{AdS}_9$-sliced 10D metric and a non-trivial profile for the type IIB axio-dilaton. The explicit IIB field \emph{Ansatz} reads
\begin{equation}
\label{AdS9_Ansatz}
\begin{split}
&ds^{2}_{10} \, = \, e^{2A(y)}L^{2}ds_{\mathrm{AdS}_{9}}^{2} \, + \, e^{2B(y)} dy^2\ , \vspace{4mm}\\
&\tau \,=\, C_{(0)}(y) \,+\, i \, e^{-\Phi(y)} \ .
\end{split}
\end{equation}
The equations of motion of type IIB supergravity in the Einstein frame simply reduce to
\begin{eqnarray}
\label{eq:eomIIBAdS9}
\hspace{-2mm}0&=&\frac{28}{L^2} e^{2(B-A)}+8\left(A''-A'B'+\frac{9}{2}(A')^2\right)+\frac{(\Phi')^2+e^{2\Phi}( C_{(0)}')^2}{4}\ , \nonumber \\
\hspace{-2mm}0&=&\left(e^{9A-B}\Phi'\right)' -e^{9A-B}e^{2\Phi}(C_{(0)}')^2 \ , \label{AdS9_EOMs}\\
\hspace{-2mm}0&=&\left(e^{9A-B}e^{2\Phi}C_{(0)}'\right)'  \ , \nonumber 
\end{eqnarray}
where $'$ denotes a derivative w.r.t. $y$. As one expects, the above second order system is subject to an additional \emph{Hamiltonian constraint} coming from the Einstein equation along $yy$ directions, which is given by
\begin{equation}\label{HCD}
\frac{36}{L^2} e^{2(B-A)}+36(A')^2-\frac{(\Phi')^2+e^{2\Phi}(C_{(0)}')^2}{4}=0 \ .
\end{equation}
This constraint is compatible with the above second order system, in the sense that its $y$ derivative identically vanishes on-shell, as a consequence of the second order field equations \eqref{AdS9_EOMs}. 

The general solution to the above field equations was discussed in \cite{Dibitetto:2026oor}, as a special case of a class of backgrounds in arbitrary dimensions found in \cite{Dibitetto:2026yft}. Reparameterizing the coordinate on the interval, we impose the $e^{B(y)}=L\,e^{A(y)} $ gauge, and the metric and axio-dilaton get a very simple form
\begin{equation}\left\{
\begin{array}{lclc}
ds_{10}^2 & = &L^2 \sin^{\frac{1}{4}}\left(8 y\right) \, \left( ds_{\text{AdS}_9}^2+dy^2\right) & ,\\[3mm]
e^{\Phi} & = &  c_1 \,\tan^{\frac{3}{2}}(4y)+c_2\, \tan^{-\frac{3}{2}}(4 y) & ,\\[3mm]
C_{(0)} & = &  \chi_0 - \sqrt{\dfrac{c_2}{c_1}}\left(c_2 +c_1\tan^{3}(4 y)\right)^{-1}& ,
\end{array}\right. \label{AdS9_SOL}
\end{equation}
where $\chi_0$ and $c_{1,\,2}>0$ are real integration constants. The 10D spacetime topology is a singular suspension of $\text{AdS}_9$ which is homeomorphic to $\mathbb{R}^{10}$ with two singularities corresponding to the endpoints of the interval. The length of the interval $I$ is fixed by the AdS radius $L$ and the coordinate range of $y \in [0,\,\pi/8] = I$.

The above solution, when Wick rotated to Euclidean signature, may be argued to possess a Euclidean wormhole interpretation, in the sense that the bulk 10D metric is a tube connecting two $\mathbb{H}^9$ boundaries separated at finite geodesic distance from each other.
As already remarked in \cite{Dibitetto:2026oor}, the 10D geometry ends up in a singularity at both boundaries of $I$. Meanwhile, at $\partial I$ the dilaton diverges. However, it was argued there that the finiteness of its on-shell Euclidean action indicates that these singularities actually get resolved into higher-dimensional flat space. Moreover, it is worth mentioning that the above solution can be obtained from a \textit{seed solution} with vanishing axion
\begin{equation}\left\{
\begin{array}{lclc}
ds_{10}^2 & = &L^2 \sin^{\frac{1}{4}}\left(8 y\right) \, \left( ds_{\text{AdS}_9}^2+dy^2\right) & ,\\[2mm]
e^{\Phi} & = & \tan^{\frac{3}{2}}(4 y) & ,\\[2mm]
C_{(0)} & = &  0 & .
\end{array}\right. \label{AdS9_C0_SOL}
\end{equation}
In this context, by using \eqref{AdS9_C0_SOL} as a seed, the general solution in \eqref{AdS9_SOL} may be generated by acting on \eqref{AdS9_C0_SOL} with a general $\mathrm{SL}(2,\mathbb{R})$ transformation, which maps solutions of the type IIB equations of motion among them. The peculiarity of \eqref{AdS9_C0_SOL} is that the dilaton only diverges at one of the endpoints of $I$, while at the other one there is a weakly coupled description in terms of $g_s$. When it comes to holographic interpretations though, the seed solution turns out to be more difficult to understand than a generic one, as it has a vanishing $F_{(1)}$ flux.

\subsection{Flux quantization and holographic quantities}
Using standard formulas \cite{Klebanov:2007ws,Cremonesi:2015bld}, we compute the would-be holographic central charge and the on-shell Euclidean action:
\begin{equation}
\label{chol}
c_{\textrm{hol}} \,=\, \frac{6}{(4\pi\ell_s)^8}\int\limits_I (L e^A)^7 e^{B}dy \ =  \frac{6}{(4\pi\ell_s/L)^8}\int\limits_0^{\pi/8} \sin(8y)dy =\frac{3}{2(4\pi\ell_s/L)^8} \,,
\end{equation}
and
\begin{equation}
    S_{E|\text{on-shell}} = - \frac{9 L^9 \text{Vol}(\mathbb{H}^9)}{8 \pi G_{10}}\left. e^{9A-B} A'\right|_{\partial I} = \frac{18 \text{Vol}(\mathbb{H}^9)}{(4\pi \ell_s/L)^8}\,.
\end{equation}
We denote by $\textrm{Vol}(\mathbb{H}^9)$ the regularized volume of 9D hyperbolic space. In both cases we obtain an $(L/\ell_s)^8$ scaling which is independent of the choice of duality frame.

In analogy with standard AdS/CFT examples, let us now discuss flux quantization when $c_2\neq 0$. Note that this is strictly necessary to have non-vanishing flux and also, as we will see later on, it turns out to be the only way to consistently impose an R7-brane monodromy. Moreover, flux quantization should really be performed at the level of the brane solutions and the following computation could also be performed at the level of the 7-brane solutions in the next section. We define $N$ via
\begin{equation}
N \  \equiv \ \frac{1}{2\pi} \int\limits_I F_{(1)} \ = \  \frac{1}{2\pi} C_{(0)}|_{\partial I} = \frac{1}{2\pi\,\sqrt{c_1 c_2}}\ .
\end{equation}
In the same spirit, we could consider the dual axion, $C_8$, in order to obtain an AdS$_9$ flux. As per the equations of motion (which lead to the third equation in \eqref{eq:eomIIBAdS9}) in the absence of three-form fluxes we have that
\begin{equation}
    d(e^{2\Phi} \star F_1) = 0\,,
\end{equation}
hence we can define an 8-form dual to $C_{(0)}$ via
\begin{equation}
   e^{2\Phi} \star F_1 = d C_{(8)} = 24 \sqrt{c_1\,c_2} L^8\text{vol}_{\text{AdS}_9} \,.
\end{equation}
We define $k$ such that $dC_{(8)} = k\, \ell_s^8\, \text{vol}_{\text{AdS}_9}$ resulting in 
\begin{equation}
\label{eq:quantL}
    \frac{L}{\ell_s} = \left( \frac{\pi\,N k}{12}\right)^{1/8}\,.
\end{equation}
These two parameters $N$ and $k$, which are candidates to be quantized numbers, fully determine the AdS scale on which depend both the holographic central charge and the on-shell action. In particular, the quantization condition $N \in \mathbb{Z}$ is inherited from reinterpreting these AdS$_9$ solutions as the near-horizon limit of the 7-brane solutions presented in the next section. Naively using the relation \eqref{eq:quantL}, the holographic central charge \eqref{chol} would scale as $c_{\text{hol}} \propto k \,N$.

\subsection{AdS$_9$ limit}
Here, we choose coordinates in a way adapted to our R7 brane solution presented later.
Consider \eqref{AdS9_SOL}, then choose $c_1=c_2=\frac{1}{2A_\tau}$ and $\chi_0=A_\tau$ for constant $A_\tau$, and change coordinates using
\begin{equation}
    \theta=8y, \quad ds_{\text{AdS}_9}^2=\frac{dR^2+ds^2_{\text{Mink}_8}}{R^2}, \quad r=\frac{R^8}{L^7}, \quad z=re^{i\theta},
\end{equation}
so that $\theta \in [0,\pi]$ which gives
\begin{equation}\left\{
\begin{array}{lclc}
ds_{10}^2 & = &\left(\frac{L\sin\theta}{r}\right)^\frac{1}{4} ds^2_{\text{Mink}_8} +\frac{L^2\sin^\frac{1}{4}\theta}{64r^2}(dr^2+r^2d\theta^2) & ,\\[3mm]
e^{\Phi} & = &\frac{\cosh \lambda}{A_\tau} & ,\\[3mm]
C_{(0)} & = &  -A_\tau \tanh \lambda&
,\\[3mm]
\lambda&=&\frac{3}{2}\arcsinh(\cot\theta). \label{ads9final}
\end{array}\right. 
\end{equation}
These changes of variables are natural as they remove the artificial eightfold periodicity and parameterises $\tau$ to lie along a geodesic via $\tau=A_\tau(i\sech \lambda-\tanh \lambda)$.

\section{Exact R7-brane solutions}

\label{section3}

In this section we propose a method to build solutions to the axio-dilaton-Einstein equations in type IIB which possess \emph{i)} the appropriate R7-monodromy, \emph{ii)} an AdS$_9$ limit and \emph{iii)} an asymptotically Minkowski$_{10}$ geometry at infinity. The method goes in three steps. First we reduce the non-linear equations of motion to a linear subsector. In this simplified subsector, using an appropriate coordinate system, we select a seed solution with the correct monodromy for the axio-dilaton and an AdS$_9$ limit. Finally, we will add another mode to the seed solution which is subleading near the AdS$_9$ limit but allows us to tune the geometry at infinity to reach a flat geometry.

\subsection{The equations of motion}
Our starting point is exactly the one in \cite{Cavusoglu:2026xiv} and we will follow the very same notation. We consider the type IIB supergravity action restricted to the sector composed by the metric and the axio-dilaton only. In this sector the 10D Einstein frame action reads
\begin{equation}\label{action_tau}
S_{10} \ = \ \frac{1}{2\kappa_{10}^2}\,\int d^{10}x \sqrt{-g_{10}}\,\left(\mathcal{R}_{10}\,-\,\frac{\partial_\mu \tau \partial^\mu \bar{\tau}}{2\mathrm{Im}(\tau)^2}\right) \ ,
\end{equation}
where the complex axio-dilaton is defined as $\tau\,\equiv\, C_{(0)} \,+\, i\,e^{-\Phi}$. By setting the 10D gravitational coupling $\kappa_{10}=1$, the general equations of motion of this system are
\begin{eqnarray}
\frac{1}{\sqrt{-g}}\partial_{\mu}\left(\sqrt{-g}\,g^{\mu\nu}\partial_{\nu}{\tau}\right) \,+\, i\,\frac{\partial_{\mu}\tau\partial^{\mu}\tau}{\mathrm{Im}(\tau)} & =  0 \ , \\
G_{\mu\nu} \,-\, \frac{1}{4\mathrm{Im}(\tau)^2} \left(2\partial_{(\mu}\tau\partial_{\nu)}\bar{\tau}-g_{\mu\nu}\partial_{\rho}\tau\partial^{\rho}\bar{\tau}\right) & =  0 \ ,
\end{eqnarray}
where $G_{\mu\nu}$ denotes the 10D Einstein tensor. Let us now specify our metric to a 7-brane \emph{Ansatz}. We split the ten-dimensional space time as a warped product of Mkw$_8$ and a Riemann surface with complex coordinate $z$. The Ansatz for the metric is of the form
\begin{equation}
 ds_{10}^2=e^{f(z,\bar{z})} ds^2_{\text{Mkw}_{8}}+e^{h(z,\bar{z})}(dz d\bar{z}) \ ,
\end{equation}
along with an axio-dilaton $\tau(z,\,\bar{z})$. From the tracelessness of the energy-momentum tensor associated with a 2D scalar field, we get the following simple differential condition for $f$
\begin{equation}
\partial \bar{\partial}f\,+\,4 |\partial f|^2 \ = \ 0 \ ,
\end{equation}
which directly implies that $e^{f} \,=\, H^{\frac{1}{4}}$, where $H$ is \emph{harmonic}, \emph{i.e.} $\partial\bar\partial H\,=\,0$.

The 10D field equations for $f(z,\bar{z})$, $h(z,\bar{z})$ and $\tau(z,\bar{z})$ are
\begin{eqnarray}
\partial\bar{\partial}h+\frac{7}{2}\partial\bar{\partial}f  =   -\dfrac{\partial\tau\bar{\partial}\bar{\tau}+\bar{\partial}\tau{\partial}{}\bar{\tau}}{4\mathrm{Im}(\tau)^2} \ ,\\
\partial^2f+\frac{1}{2}(\partial f)^2-\partial f\partial h =  -\dfrac{\partial\tau{\partial}\bar{\tau}}{8\mathrm{Im}(\tau)^2} \ ,\\
\bar{\partial}^2f+\frac{1}{2}(\bar{\partial} f)^2-\bar{\partial} f\bar{\partial} h  =  -\dfrac{\bar{\partial}\tau{\bar{\partial}}\bar{\tau}}{8\mathrm{Im}(\tau)^2} \ , \\
\partial\bar{\partial}\tau+2\left(\partial f\bar{\partial}{\tau}+\bar{\partial} f{\partial}{\tau}\right)  =   -i\dfrac{\partial\tau\bar{\partial}\tau}{\mathrm{Im}(\tau)} \ . \label{eom}
\end{eqnarray}
 In \cite{Cavusoglu:2026xiv} it was already noted that the above PDE's are not all independent of one another. This is best seen by formulating them in terms of a function $h_0$ defined through
\begin{equation}\label{def_h0}
h \ = \ h_0 \,+\, \frac{9}{2}f \,+\, \log\left(\partial f\bar{\partial}f\right) \ .
\end{equation}
In particular, the second order Einstein equations, follow from
\begin{equation}\label{eom_h0}
\partial f \partial h_0 \,=\, \frac{\partial \tau \partial \bar{\tau}}{8\mathrm{Im}(\tau)^2} \ , \quad \bar{\partial} f \bar{\partial} h_0 \,=\, \frac{\bar{\partial} \tau \bar{\partial} \bar{\tau}}{8\mathrm{Im}(\tau)^2} \ ,
\end{equation}
by taking (anti-)holomorphic derivatives thereof. These conditions are still to be supplemented with the harmonicity constraint for $e^{4f}$, and the e.o.m. of $\tau$.

It is possible to use holomorphic change of coordinates to simplify these equations. To do so, we introduce $J(z)$ as a holomorphic function with real part $H$ and imaginary part $I$ (i.e. $I$ is the harmonic conjugate to $H$) such that
\begin{equation}  J(z)=H(z,\bar{z})+iI(z,\bar{z}), \quad \bar{\partial}J=0.
\end{equation}
Further assuming that $J$ is meromorphic, we can locally use it as the coordinate on the Riemann surface. The equations of motion further simplify to
\begin{equation}
\text{Im}(\tau)\left(\tau_{HH}+\tau_{II}+\frac{1}{H}\tau_H\right)=-i(\tau_H^2+\tau_I^2). \label{ernsteq}
\end{equation}
This is a case of a very well-studied equation known as the axisymmetric Ernst equation which is in fact integrable (see, for example, \cite{ernstsols}).

Finally, $h_0$ can be obtained from the remaining equations of motion \eqref{eom_h0} which turn into
\begin{equation}
    (h_0)_H=\frac{H}{2\text{Im}(\tau)^2}(|\tau_H|^2-|\tau_I|^2), \quad(h_0)_I=\frac{H}{2\text{Im}(\tau)^2}(\bar{\tau}_I\tau_H+\tau_I\bar{\tau}_H). \label{h0}
\end{equation}
Note that the integrability condition for $h_0$ is exactly the equation of motion for $\tau$.

The 10D metric can be written as
\begin{equation}
 ds_{10}^2=H^\frac{1}{4} ds^2_{\text{Mkw}_{8}}+\frac{e^{h_0}}{64}H^{-\frac{7}{8}}(dH^2+dI^2)  \,,
\end{equation}
where $H$ is a \emph{positive} harmonic function.

\subsubsection{Geodesic subsector}

The equation determining $\tau$ \eqref{ernsteq} has a very general solution, discussed in, for example, \cite{ernstsols}. However, we shall focus on a subset of these solutions for which a closed form can be easily written down. In particular, we can view this Ernst equation as a harmonic map into $\mathrm{SL}(2,\mathbb{R})/\mathrm{SO}(2)$ and take $\tau$ to lie on a geodesic of the target space as in \cite{plebanski}. This reduces to the \emph{Ansatz}
\begin{equation}
    \tau=A( i \sech \lambda - \tanh \lambda),
\end{equation}
where $\lambda(H,\,I)$ is a real function. Note that, up to an $\mathrm{SL}(2,\,\mathbb{R})$ transformation, we can always go to a frame where $\tau = i e^{-\lambda}$. Using this first restriction, the \eqref{ernsteq} simplifies significantly to a \emph{linear} PDE for $\lambda$:
\begin{equation}
    \text{Im}(\tau)\tau'(\lambda)\left(\lambda_{HH}+\lambda_{II}+\frac{1}{H}\lambda_H\right)=0. 
\end{equation}
Assuming $\text{Im}(\tau)\neq0$ and $\tau'(\lambda)\neq 0$, this results in the standard axisymmetric Laplace equation for $\lambda(H,I)$
\begin{equation}
    \lambda_{HH}+\lambda_{II}+\frac{1}{H}\lambda_H=0.
    \label{lap}
\end{equation}
Furthermore, the equations \eqref{h0} determining $h_0$ reduce to a simple integration problem:
\begin{equation}
    (h_0)_H+i(h_0)_I
    =
    \frac{H}{2}
    \left(\lambda_H+i\lambda_I\right)^2\,.
\end{equation}
Note that the integrability condition for $h_0$ is a consequence of the equations of motion for $\lambda$.

\subsubsection{Mode expansion and separation of variables}

Importantly for us, the equation \eqref{lap} is a linear PDE and thus a solution can be built from a sum of modes obtained by a separation of variable technique. Here we will rely on two coordinate systems.

\paragraph{Cylindrical coordinates and the AdS$_9$ solution}

The first one relies on cylindrical coordinates. Using $H= r\sin\theta$ and $I=r\cos\theta$ with $r>0$ and $\theta \in (0,\,\pi)$, the equations for $\lambda$ and $h_0$ reduce to
\begin{equation}
\begin{split}
    &\partial_r (r^2 \partial_r \lambda) + \frac{1}{\sin\theta} \partial_\theta (\sin\theta \,\partial_\theta \lambda) = 0\,,\\
    &\partial_r h_0 - \frac{i}{r} \partial_\theta h_0 = i \frac{r \sin\theta}{2} e^{-i\theta} (\lambda_r - \frac{i}{r} \lambda_\theta)^2\,.
    \end{split}
\end{equation}
Using separation of variables, the various modes for $\lambda$ are of the form
\begin{equation}
    \lambda_n = (a_n r^n + b_n r^{-(n+1)}) (c_n P_n(\cos\theta) + d_n Q_n(\cos\theta))
\end{equation}
where $n>-1/2$ is a real number and $P$ and $Q$ are the Legendre functions.

Using these coordinates, the AdS$_9$ solution simply corresponds to the $r$-independent solutions, i.e. to the single mode 
\begin{equation}
\begin{split}
    \lambda_{\text{AdS}_9} &= \frac{3}{2} Q_0(\cos\theta) + \text{const.} = \frac{3}{4}\log\left(\frac{1+\cos\theta}{1-\cos\theta} \right)\,\\
    &=-\frac{3}{2}\log\tan\left(\frac{\theta}{2}\right)\,.
\end{split}
\end{equation}
up to a rescaling $\theta \rightarrow 8 \theta$ and $r\rightarrow r^8$ yielding back the solution \eqref{AdS9_SOL}.

\paragraph{Toroidal coordinates and the R7-brane solution}

The second system of coordinates at play here are the toroidal coordinate $(\eta,\,\Theta)$ 
\begin{equation}
    e^{\eta- i \Theta} = \frac{J+1}{J-1}\ ,
\end{equation}
for which it follows that $H$ and $I$ as expressed in terms of $\eta$ and $\Theta$ as
\begin{equation}
   H=\frac{\sinh \eta}{\cosh \eta-\cos \Theta}, \quad  I=\frac{\sin \Theta}{\cosh \eta-\cos \Theta}. \label{toroidalcoordinates}
\end{equation}
This change of coordinates maps the half-plane $H>0$ to the exterior of the unit disk in the complex plane, \emph{i.e.} $\eta >0$ and $\Theta \in [0,\,2\pi)$, importantly, the point $J=1$ corresponds to the point at infinity in this chart. This leaves us with a coordinate chart with the topology of an open cylinder, which is more appropriate to discuss various limiting behaviours and monodromy (\emph{i.e.}, it is not simply connected).

The corresponding mode expansion \cite{toroidal} is
\begin{equation}
\begin{split}
    \lambda_n=&\sqrt{D} \left(A_nP_{n-\frac{1}{2}}(\cosh \eta)+B_nQ_{n-\frac{1}{2}}(\cosh \eta)\right)\left(C_n\cos n \Theta + D_n \sin n\Theta\right)\,,
    \label{eq:modesToroidal}
\end{split}
\end{equation}
where
\begin{equation}
    \begin{split}
    D=& \cosh \eta-\cos \Theta\,,  \label{modeexpansion}
    \end{split}
\end{equation}
and $P_{n}$ and $Q_n$ are, once again, Legendre functions. In this coordinate system, the AdS$_9$ solution corresponds to 
\begin{equation}
    \lambda_{\text{AdS}_9}=\frac{3}{2}\arcsinh\left(\frac{\sin \Theta}{\sinh \eta}\right)\,,
    \label{eq:lambdaAdS9Toroidal}
\end{equation}
and $h_0$ must satisfy
\begin{equation}
\partial_\eta h_0+i\partial_\Theta h_0
=
\frac{\sinh\eta}{2(\cosh\eta-\cos\Theta)}
\left(1-\cosh(\eta-i\Theta)\right)
\left(\partial_\eta\lambda+i\partial_\Theta\lambda\right)^2\,.
\end{equation}
\subsection{An R7-brane solution}

Now that we have simplified the equations of motion, we will build the R7-brane solution with the appropriate properties.

\paragraph{Seed solution and R7-brane monodromy}

First we would like to impose the correct monodromy for the R7 brane. In term of $\lambda$, this corresponds to imposing the presence of a branch cut where $\lambda \rightarrow - \lambda$. Using the toroidal coordinates, imposing $\lambda(\eta,\,\Theta) = - \lambda(\eta,\,\Theta + 2\pi)$ implies that only the half-integer modes $n \in \mathbb{Z}+\tfrac{1}{2}$ in \eqref{eq:modesToroidal}.

Remarkably, the following identities hold (see Appendix.A. of \cite{andrews})
\begin{equation}
\begin{split}
\lambda_{0}
=&\frac{3}{2}\sqrt{2D}\sum_{n=0}^{+\infty} Q_n(\cosh \eta) \sin\left( \left(n+\frac{1}{2}\right)\Theta\right) = \frac{3}{2} \arcsinh \left(\frac{\sin \tfrac{\Theta}{2}}{\sinh\tfrac{\eta}{2}}\right)\label{resummation}
\end{split}
\end{equation}
and does indeed solve equation \eqref{lap}\footnote{At first glance, this solution resembles a simple holomorphic change of coordinate $z= e^{\eta - i\theta} \rightarrow z' = e^{\tfrac{1}{2}(\eta - i\theta)}$ when compared with $\lambda_{\text{AdS}_9}$. However, this solution does solve the equations of motion for $\lambda$ with respect to the original $H$ harmonic function \eqref{toroidalcoordinates} and not the rescaled function ${H' = \tfrac{\sinh\tfrac{\eta}{2}}{\cosh\tfrac{\eta}{2}-\cos\tfrac{\Theta}{2}}}$. }. This solution has a different mode expansion as $\lambda_{\text{AdS}_9}$ and is a genuinely different solution. From $\lambda_0$ we compute that $h_0$ must satisfy the equations
\begin{equation}
     \partial_\eta h_0 =\frac{9}{16}\coth\frac{\eta}{2}, \quad \partial_{\Theta}h_0=0.
\end{equation}
These are solved explicitly by
\begin{equation}
    e^{h_0}=c_0\left(2 \sinh \frac{\eta}{2}\right)^\frac{9}{8},
\end{equation}
for constant $c_0$. Since this does not depend on $\Theta$, it is smoothly defined across the angular coordinate. The final form of the metric is
\begin{equation}
\begin{split}
 ds_{10}^2=&\left(\frac{\sinh \eta}{D} \right)^\frac{1}{4}ds^2_{\text{Mkw}_{8}}+ c_0 \frac{(2 \sinh \tfrac{\eta}{2})^{9/8}}{64}\frac{1}{(\sinh \eta/D)^\frac{7}{8}}\frac{1}{D^2}(d\eta^2+d\Theta^2),  \label{sthetametric}\\
 D=&\cosh \eta-\cos \Theta.
 \end{split}
\end{equation}
The monodromy is given by $\Theta\rightarrow \Theta+2\pi$ under which
\begin{equation}
    \lambda\rightarrow - \lambda, \quad h_0\rightarrow h_0, \quad H\rightarrow H.
\end{equation}

\paragraph{The AdS$_9$ limit}
As $\eta\rightarrow 0$ we observe that the $\lambda$ diverges to $\pm \infty$ depending on the sign of $\Theta$.
Following our various changes of coordinates, this limit corresponds to the boundary value of the axio-dilaton in the AdS$_9$ case which in polar coordinates corresponded to $\theta \rightarrow \pm \pi/2$. We show now how to recover the AdS$_9$ geometry in the limit where $\eta \rightarrow 0$ while keeping the ratio $\tfrac{\sinh\tfrac{\eta}{2}}{\sin \tfrac{\Theta}{2}}$ fixed (implying $\Theta\rightarrow 0$). This is simply analogous to taking a radial coordinate to zero in polar coordinates. To take this limit, we define
\begin{equation}
    x = \sinh\tfrac{\eta}{2}\ , \hspace{1cm} y = \sin \tfrac{\Theta}{2}\ ,
\end{equation}
and observe that $D = 2 (x^2 + y^2)$. Going to polar coordinates $x = r \cos \theta$ and $y = r \sin \theta$, we can expand near $r \rightarrow 0^+$ and obtain
\begin{equation}
    ds^2_{10} \rightarrow \cos^\frac{1}{4} \theta \left(r^{-1/4}ds^2_{\text{Mkw}_8} + \kappa^2 \left(\frac{dr^2}{r^2} +  d\theta^2\right) \right)\,.
\end{equation}
We recognise the AdS$_9 \ltimes I$ solution of the previous section where $\kappa^2$ is a numerical factor depending on $c_0$ and which fixes the AdS$_9$ scale.

\paragraph{Tuning the asymptotics}

Finally, we want to study the limit where $\eta \rightarrow +\infty$. In that limit $\lambda \rightarrow 0$ and the axio-dilaton is constant. However, the metric expands as
\begin{equation}
    ds_{10}^2 \rightarrow ds^2_{\text{Mkw}_8} + e^{-\tfrac{23}{16}\eta } (d\eta^2 + d\Theta^2)\ ,
\end{equation}
which collapses the $(\eta,\,\Theta)$ fibre and results in a curvature singularity as $\eta\rightarrow +\infty$. To obtain a proper flat limit as $\eta \rightarrow +\infty$, the prefactor of the transverse $(\eta,\,\Theta)$ metric must asymptote $e^{2\eta} (d\eta^2  + d\Theta^2)$.

A way to solve this is to add an extra mode to $\lambda$ which is subleading in the AdS$_9$ limit and which provides the appropriate correction to $h_0$ in order to reach a flat limit at infinity. 

We will add to $\lambda_0$ the lowest $Q$ mode preserving the monodromy:
\begin{equation}
    \lambda_\infty=\alpha\sqrt{2D}Q_0(\cosh \eta) \sin \frac{\Theta}{2}=\alpha\sqrt{2D}\log\left(\coth \frac{\eta}{2}\right)\sin \frac{\Theta}{2}\ ,
\end{equation}
for $\alpha$ a constant to be fixed. Taking the same AdS$_9$ limit as before, this mode asymptotes
\begin{equation}
    \lambda_{\infty} \rightarrow -2\alpha r^2 \log(r \cos\theta) \sin \theta\ ,
\end{equation}
which is sub-leading compared to $\lambda_0$ as $r\rightarrow 0$ (as it should).

Now, in order to obtain the proper Mink$_{10}$ asymptote, we must require $h_0 \rightarrow 4 \eta$ as $\eta \rightarrow +\infty$. We compute that
\begin{equation}
    \lambda_0+\lambda_\infty \rightarrow (3+2\alpha)e^{-\frac{1}{2}\eta}\sin \frac{\Theta}{2}+\mathcal{O}(e^{-\frac{3}{2}\eta})\ .
\end{equation}
Solving asymptotically for $h_0$ we obtain that as $\eta\rightarrow \infty$,
\begin{equation}
    e^{h_0}\rightarrow c_02^{-\frac{\alpha^2}{2}}e^{\frac{3}{8}\alpha(\alpha+2)}e^{\frac{(3+2\alpha)^2}{16}\eta}(1+\mathcal{O}(e^{-\eta})). \label{hlim}
\end{equation}
This leads to a metric with asymptotic behaviour:
\begin{equation}
    ds_{10}^2\rightarrow ds^2_{\text{Mkw}_8}+C_1e^{\left(\frac{(3+2\alpha)^2}{16}-2\right)\eta}(d\eta^2+d\Theta^2),
\end{equation}
for a constant $C_1$. Choosing either $\alpha = \tfrac{5}{2}$ or $\alpha =- \tfrac{11}{2}$ yields the correct asymptotic behaviour to Minkowski space.

From this asymptotic behaviour of the metric, we can see that for $\alpha=\frac{5}{2}$ or $\alpha=-\frac{11}{2}$, the metric is asymptotically flat, which is our main case of interest. We choose $\alpha=\frac{5}{2}$ but either could be chosen.
However, it has also been of interest to consider conical deficits in these solutions \cite{Dierigl:2022reg}. We can see that for $-\frac{11}{2}<\alpha<-\frac{3}{2}-2\sqrt{2}$ and for $-\frac{3}{2}+2\sqrt{2}<\alpha<\frac{5}{2}$, the metric is asymptotically conical with deficit angle
\begin{equation}
    \delta(\alpha)=2\pi\left(1-\left|\frac{(3+2\alpha)^2}{32}-1\right|\right). \label{angle}
\end{equation}
Note that taking $\alpha<-\frac{11}{2}$ or $\alpha>\frac{5}{2}$ leads to an asymptotically conical metric with $\delta<0$ and $|\delta|$ now defining the excess angle.
For $\alpha=-\frac{3}{2}\pm 2 \sqrt{2}$, the metric is asymptotically cylindrical.
For $-\frac{3}{2}-2\sqrt{2}<\alpha<-\frac{3}{2}+2\sqrt{2}$, there is a curvature singularity.

\subsection{Summary of the R7-brane solution}

To summarise, with coordinates defined as
\begin{equation}
\quad e^{-\eta+i\Theta}=\frac{J-1}{J+1}\,,
\end{equation}
the R7 brane solution is
\begin{equation}
\begin{split}
 ds_{10}^2=&\left(\frac{\sinh \eta}{D} \right)^\frac{1}{4}ds^2_{\text{Mkw}_{8}}+\frac{e^{h_0(\eta,\Theta)}}{64}\frac{1}{D^\frac{9}{8}(\sinh \eta)^\frac{7}{8}}(d\eta^2+d\Theta^2),  \label{sthetametric}\\
 D=&\cosh \eta-\cos \Theta.
 \end{split}
\end{equation}
where $\lambda$ is
\begin{equation}
    \lambda=\lambda_0+\lambda_{\infty}=\frac{3}{2}\arcsinh\left(\frac{\sin \frac{\Theta}{2}}{\sinh \frac{\eta}{2}}\right)+\frac{5}{\sqrt{2}}\sqrt{D}\log\left(\coth \frac{\eta}{2}\right)\sin \frac{\Theta}{2}, \label{lambdafinal}
\end{equation}
and as a consequence, the axio-dilaton reads
\begin{equation}
    \tau=A(i\sech \lambda - \tanh \lambda),
\end{equation}
while the complete expression for $h_0$ is given by
\begin{equation}
\begin{split}
    h_0(\eta,\Theta)=&\log C + \frac{9}{8}\log \left(\sinh \frac{\eta}{2}\right)\\
    +&\frac{3\alpha}{4}\left[-\cos\Theta+\log(\sinh \eta)+((1+\cosh \eta)\cos\Theta-1)\log\left(\coth \frac{\eta}{2}\right)\right]\\
    +&\frac{\alpha^2}{8}\left[-4\cos\Theta+\cos(2\Theta)+4\log\left(\cosh \frac{\eta}{2}\right)\right.\\
    &\left.+(6-4(1+\cosh \eta)\cos \Theta+2\cosh \eta \cos (2\Theta))\log\left(\tanh \frac{\eta}{2}\right)\right.\\
    +&(\cosh \eta-1) ((1+\cosh \eta)\cos(2\Theta)-2)\left(\log\left(\coth \frac{\eta}{2}\right)\right)^2\\
    +&\left.2\,\mathrm{Li}_2\left(\sech^2 \frac{\eta}{2}\right)\right],
\end{split} \label{h0final}
\end{equation}
with $\alpha=\frac{5}{2}$.

\begin{figure}
\begin{center}
\begin{tikzpicture}[
    >=Stealth,
    thick,
    every node/.style={font=\large}
]

\fill[black](1.15,0.0) circle(0.1);

% The n=0 disk
\fill[gray!25] (0,0) circle (1.15);
\draw (0,0) circle (1.15);

% Branch cut
\draw[
    decorate,
    decoration={
        snake,
        amplitude=0.10cm,
        segment length=0.28cm
    }
]
(-1.15,0) -- (-4.0,0);

\node at (-2.55,0.55) {\small Branch cut};

% eta arrow (radial, pointing outward from the eta=0 boundary)
\draw[->] (76:1.15) -- (76:1.65);
\node at (0.55,1.75) {\small$\eta$};
\node at (-95:0) {\small $\eta < 0$};

% Angular direction (tangential arc on the boundary)
\draw[->] (20:1.35) arc (20:70:1.35);
\node at (45:1.65) {$\Theta$};

% Coordinate transformation with half-square bracket
\node (zdef) at (-5.2,2.) {\small$Z=e^{\eta-i\Theta}$};
\draw ([xshift=-3mm,yshift=-2mm]zdef.south west)
   -- ([xshift=2mm,yshift=-2mm]zdef.south east)
   -- ([xshift=2mm,yshift=3mm]zdef.north east);

% AdS region label
\node at (2.4,-0.2) {\small AdS$_9$ region};

\end{tikzpicture}
\end{center}
\caption{\it Diagram of the solution on the $Z = e^{\eta - i \Theta}$ plane.}
\end{figure}

This solution presents both the correct monodromy as $\Theta\rightarrow \Theta + 2 \pi$ (the metric is unchanged under $\Theta\rightarrow \Theta+2\pi$ and $\lambda(\eta,\Theta)=-\lambda(\eta,\Theta+2\pi)$, which is equivalent to $\tau \rightarrow -\bar{\tau}$) and the appropriate AdS$_9$ and Mink$_{10}$ asymptotics. In particular, in the limit $(\eta,\Theta)\rightarrow (0,0)$ with $\frac{\sinh \frac{\eta}{2}}{\sin \frac{\Theta}{2}}$ fixed, we recover the AdS$_9\ltimes I$ solution.

The defect for which the solution becomes singular is given by $\eta<0$ and ($\eta=0, \Theta \neq 0$). This is to be interpreted as a finite core for which nonperturbative effects become important, as discussed in \cite{Cavusoglu:2026xiv}.

\paragraph{Perturbative expansion} Finally, in order to compare with the perturbative expansion presented in \cite{Cavusoglu:2026xiv}, we can simply perform the appropriate holomorphic coordinate change:
\begin{equation}
    J=1+\frac{2a}{z}\ .
\end{equation}
The metric is then given by
\begin{equation}
\begin{split}
 ds_{10}^2=&H^\frac{1}{4}ds^2_{\text{Mkw}_{8}}+H^{-\frac{7}{8}}\frac{a^2}{16(z\bar{z})^2}e^{h_0(\eta,\Theta)}dzd\bar{z}\ ,
 \end{split}
\end{equation}
with
\begin{equation}
    H=1+a\left(\frac{1}{z}+\frac{1}{\bar{z}}\right).
\end{equation}

Noting again that $\tau=A(i\sech \lambda - \tanh \lambda)$, $h_0$ is given by \eqref{h0final} and $\lambda$ is given by \eqref{lambdafinal}, which can both be converted from $(\eta,\Theta)$ to $(z,\bar{z})$ coordinates via
\begin{equation}
    e^{-\eta+i\Theta}=\frac{a}{z+a}.
\end{equation}

We can choose the integration constant $c_0$ in $h_0$ such that all constant terms sum to zero in $h$. After doing this, we get the perturbative expansion in $|z|^{-1}$ as $|z|\rightarrow \infty$ for $h$
\begin{equation}
    h=\frac{49a}{8}\left(\frac{1}{z}+\frac{1}{\bar{z}}\right)-\frac{8a}{\sqrt{z\bar{z}}}-\frac{57a^2}{8}\left(\frac{1}{z^2}+\frac{1}{\bar{z}^2}\right)-\frac{25a^2}{8z \bar z}+\frac{22a^2}{3\sqrt{z\bar{z}}}\left(\frac{1}{z}+\frac{1}{\bar{z}}\right)+\mathcal{O}(|z|^{-3}).
\end{equation}
In particular, all functions can be expanded similarly and compared to the coefficients in \cite{Cavusoglu:2026xiv}. There is a freedom to add any finite number of higher modes to $\lambda$ in \ref{lambdafinal} of the form
\begin{equation}
C_i\sqrt{D}Q_i(\cosh \eta)\sin \left(\left(i+\frac{1}{2}\right)\right), \quad i=1,2,...,
\end{equation}
for constants $C_i$, which will not change the AdS$_9$ limit or asymptotic behaviour. It would be interesting to consider how the more general Ernst solutions could result in more general perturbative expansions.

\section{R7-brane solution with more conventional near-horizon limit}
\label{addedsection}

Although the solution presented in the last section tends to AdS$_9\ltimes I$ near $J\rightarrow \infty$, the point it approaches corresponds to the conformal boundary of AdS$_9$.\footnote{We thank Mirjam Cveti{\v c} and Jeffrey Kuntz for pointing this out.} We can ask whether there is a more conventional near-horizon limit, which approaches the Poincar\'{e} horizon which lies at $J\rightarrow 0$ or, equivalently, $(\eta,\Theta)=(0,\pi)$. In this section, we present such a solution, which is written explicitly up to an integral.

\subsection{Deriving the solution}

Consider the general mode expansion (with the correct monodromy) of the Laplace equation \ref{lap}
\begin{equation}
\lambda=\sum_{n=0}^\infty \sqrt{2D}Q_{n}(\cosh \eta)\left[a_n\cos \left( n+\frac{1}{2}\right)\Theta+b_n\sin  \left(n+\frac{1}{2}\right)\Theta\right],
\end{equation}
only including $Q_n$ modes and not $P_n$ modes so they decay at $\eta\rightarrow \infty$.

Furthermore, as $\eta \rightarrow 0$ and $\Theta \rightarrow \pi$, taking $\eta \ll |\pi-\Theta|$ (we will later check the near-horizon limit without this assumption),
\begin{equation}
\lambda_{\text{AdS}_9}=\frac{3}{2}\arcsinh\left(\frac{\sin\Theta}{\sinh \eta}\right)\rightarrow \frac{3}{2}\text{sgn}(\pi-\Theta)\log\left(\frac{2|\pi-\Theta|}{\eta}\right),
\end{equation}
with subleading corrections.

The general mode expansion in the same limit goes as
\begin{equation}
    Q_n(\cosh \eta)\rightarrow  \log \left(\frac{2}{\eta}\right)-\sum_{k=1}^{n} \frac{1}{k}+\mathcal{O}(\eta^2 \log \eta).
\end{equation}

We can make the (half-integer) Fourier expansion (choosing $\cos$ modes since $\lambda_{\text{AdS}_9}$ is odd around $\Theta=\pi$),
\begin{equation}
\text{sgn}(\pi-\Theta)=\frac{1}{\pi}\sum_{n=0}^\infty \frac{2(-1)^n}{(n+\frac{1}{2})}\cos\left(n+\frac{1}{2}\right)\Theta.
\end{equation}

Therefore, we choose
\begin{equation}
    a_n=\frac{3(-1)^n}{(2n+1)\pi}.
\end{equation}

So the candidate for $\lambda$ to give the correct near-horizon limit is
\begin{equation}
    \lambda=\sum_{n=0}^\infty \frac{3(-1)^n}{(2n+1)\pi}\sqrt{2D}Q_n(\cosh \eta)\cos\left(n+\frac{1}{2}\right)\Theta. \label{lambdacand}
\end{equation}

The factor of $(-1)^n$ can be absorbed using $(-1)^n \cos \left(n+\frac{1}{2}\right)\Theta=\sin\left(\left(n+\frac{1}{2}\right)\left(\pi-\Theta\right)\right)$ and $(2n+1)$ suggests differentiating with respect to $\Theta$. Therefore, we can write

\begin{equation}
\begin{split}
    -\partial_\Theta\left(\frac{\lambda}{\sqrt{2D}}\right)=&\sum_{n=0}^\infty \frac{3}{2\pi}Q_n(\cosh \eta) \cos\left(\left(n+\frac{1}{2}\right)\left(\pi-\Theta\right)\right).
\end{split}
\end{equation}

Now we can use, for example, Eq.A6. of \cite{andrews} to get, after defining $\widetilde{D}=\cosh \eta+\cos \Theta$, 
\begin{equation}
   -\partial_\Theta\left(\frac{\lambda}{\sqrt{2D}}\right)=\frac{3}{2\pi}\frac{1}{\sqrt{2\widetilde{D}}}\arctan\left(\frac{2\sin \frac{\Theta}{2}}{\sqrt{2{\widetilde{D}}}}{}\right),
\end{equation}
or, equivelently,
\begin{equation}
    \lambda=\sqrt{D}\int_\Theta^\pi d \Theta' \frac{3}{2\pi \sqrt{\widetilde{D}}}\arctan\left(\frac{\sqrt{2}\sin \frac{\Theta'}{2}}{\sqrt{{\widetilde{D}}}}\right).
\end{equation}

Defining
\begin{equation}
    x=\sinh \frac{\eta}{2}, \quad y=\sin \frac{\pi-\Theta}{2},
\end{equation}
and going to polar coordinates $x=r\sin \theta, y=r\cos \theta$, 
\begin{equation}
   \partial_y\left(\frac{\lambda}{\sqrt{2D}}\right)=\frac{3}{2\pi}\frac{1}{\sqrt{x^2+y^2}\sqrt{1-y^2}}\arctan\left(\frac{\sqrt{1-y^2}}{{\sqrt{x^2+y^2}}}\right).
\end{equation}
As $r\rightarrow 0^+$,
\begin{equation}
   \partial_y\left(\frac{\lambda}{\sqrt{2D}}\right)\rightarrow\frac{3}{4}\frac{1}{\sqrt{x^2+y^2}}.
\end{equation}
Integrating at fixed $x$ (and noting that $D\rightarrow 2$) gives
\begin{equation}
    \lambda \rightarrow \frac{3}{2}\arcsinh \left(\frac{y}{x}\right) + \mathcal{O}(r),
\end{equation}
as required to give the AdS$_9\ltimes I$ near-horizon.

Now, consider the asymptotics as $\eta\rightarrow \infty$. $\lambda$ in \ref{lambdacand} goes like
\begin{equation}
    \lambda =\frac{6}{\pi}e^{-\frac{\eta}{2}}\cos \frac {\Theta} {2} -\frac{1}{3\pi}e^{-\frac{3}{2}\eta}\left(9\cos \frac{\Theta}{2}+13 \cos \frac{3\Theta}{2}\right)+\mathcal{O}(e^{-\frac{5}{2}\eta}).
\end{equation}

To give the correct asymptotics, we want
\begin{equation}
    \lambda\sim e^{-\frac{1}{2}\eta} \sin \frac{\Theta}{2} + \mathcal{O}(e^{-\frac{3}{2}\eta}),
\end{equation}
so we first subtract the $n=0$ mode from \ref{lambdacand} then want to add in modes to give the correct asymptotic behaviour without spoiling the near-horizon behaviour. Note that removing the $n=0$ mode does not change the near-horizon behaviour.

To do this we add to $\lambda$
\begin{equation}
     \beta\sqrt{2D}\left(Q_0(\cosh \eta)\sin \frac{\Theta}{2}+3Q_1(\cosh \eta)\sin \frac{3\Theta}{2}+2Q_2(\cosh \eta)\sin \frac{5\Theta}{2}\right),
\end{equation}
which as $\eta \rightarrow 0$ and $\Theta \rightarrow \pi$, tends to
\begin{equation}
    2\beta\left(\left(\log \frac{2}{\eta}\right)-3\left(\log \frac{2}{\eta}-1\right)+2\left(\log \frac{2}{\eta}-\frac{3}{2}\right)\right)+\mathcal{O}(\eta^2 \log \eta).
\end{equation}
and as $\eta \rightarrow \infty$, tends to
\begin{equation}
    2\beta e^{-\frac{\eta}{2}} \sin \frac{\Theta}{2}+\mathcal{O}(e^{-\frac{3}{2}\eta}).
\end{equation}

Therefore, our full expression for $\lambda$ is 
\begin{equation}
\begin{split}
    \lambda=&\sum_{n=1}^\infty \frac{3(-1)^n}{(2n+1)\pi}\sqrt{2D}Q_n(\cosh \eta)\cos\left(n+\frac{1}{2}\right)\Theta\\
   &+\beta\sqrt{2D}\left(Q_0(\cosh \eta)\sin \frac{\Theta}{2}+3Q_1(\cosh \eta)\sin \frac{3\Theta}{2}+2Q_2(\cosh \eta)\sin \frac{5\Theta}{2}\right).
    \end{split} \label{lambdalast}
\end{equation}
which has the correct AdS$_9$ limit as $(\eta,\Theta)\rightarrow (0,\pi)$. This can be seen since
\begin{equation}
    \lambda \rightarrow \lambda_{\text{AdS}_9}+\mathcal{O}(r \log r),
\end{equation}
for $r\rightarrow 0^+$ with fixed $\theta$. The corresponding metric then tends to Minkowski space for $\beta=4$. As before, the metric can be made asymptotically conical with arbitrary deficit angle for different values of $\beta$.

\subsection{R7-brane with AdS$_9$ near-horizon}
To summarise, with coordinates defined as
\begin{equation}
\quad e^{-\eta+i\Theta}=\frac{J-1}{J+1}\,,
\end{equation}
with the axio-dilaton defined by
\begin{equation}
    \tau=A(i\sech \lambda-\tanh \lambda),
\end{equation}
we have
\begin{equation}
\begin{split}
    \lambda=&\sum_{n=1}^\infty \frac{3(-1)^n}{(2n+1)\pi}\sqrt{2D}Q_n(\cosh \eta)\cos\left(n+\frac{1}{2}\right)\Theta\\
   &+\beta\sqrt{2D}\left(Q_0(\cosh \eta)\sin \frac{\Theta}{2}+3Q_1(\cosh \eta)\sin \frac{3\Theta}{2}+2Q_2(\cosh \eta)\sin \frac{5\Theta}{2}\right). \label{secondlambda}
    \end{split}
\end{equation}
The metric is
\begin{equation}
\begin{split}
 ds_{10}^2=&\left(\frac{\sinh \eta}{D} \right)^\frac{1}{4}ds^2_{\text{Mkw}_{8}}+\frac{e^{h_0(\eta,\Theta)}}{64}\frac{1}{D^\frac{9}{8}(\sinh \eta)^\frac{7}{8}}(d\eta^2+d\Theta^2),  \\
 D=&\cosh \eta-\cos \Theta.
 \end{split}
\end{equation}
where $h_0$ is given by
\begin{equation}
\partial_\eta h_0+i\partial_\Theta h_0
=
\frac{\sinh\eta}{2(\cosh\eta-\cos\Theta)}
\left(1-\cosh(\eta-i\Theta)\right)
\left(\partial_\eta\lambda+i\partial_\Theta\lambda\right)^2\,. \label{secondh0}
\end{equation}
In other words, $\partial_\eta h_0$ and $\partial_\Theta h_0$ can be written analytically using the expression for $\lambda$. However, integrating to get $h_0$ explicitly seems difficult.
This gives an AdS$_9$ near-horizon, Minkowski asymptotics and the correct monodromy outside the core.

\section{The 12D geometric picture}\label{section4}
The type IIB backgrounds discussed in this paper only involve the 10D metric and the axio-dilaton and hence they may be seen as classical extrema of the action \eqref{action_tau}. The axio-dilaton $\tau$ may be interpreted as the complex structure of an auxiliary $\mathbb{T}^2$ torus (or some other 2D space as we will mention). As a consequence, it becomes very natural to view our solutions as F-theoretic configurations, where the axio-dilaton profile over 10D spacetime gets translated into the geometric data that specify how the $\mathbb{T}^2$  is fibered over spacetime. The 12D geometry of the total space of this fibration may be written as
\begin{equation}
ds_{12}^2 \,=\, ds_{10}^2 \,+\, \frac{1}{\mathrm{Im}(\tau)}\left(|\tau|^2dx^2+dy^2-2\mathrm{Re}(\tau)dx dy\right) \ .
\end{equation}
The 12D pure Einstein-Hilbert action reduced on $\mathbb{T}^2$ according to the above KK \emph{Ansatz} exactly yields \eqref{action_tau}, which is nothing but the type IIB supergravity action restricted to the metric and axio-dilaton subsector.

The corresponding 12D geometric picture, though useful in some cases, is merely to be viewed here as a book-keeping device and it has no ambition to be a physical effective description valid in a certain regime. The reasons why this cannot be the case are various, starting from the fact that there is no supergravity theory with 32 real supercharges in spacetime signature $(1,11)$, and continuing with the fact that the volume of the auxiliary torus is frozen to one in a non-dynamical way. We refer to the recent work of \cite{Cheng:2025efp} for an overview of the status of the 12D geometric description for type IIB string theory.

The 12D geometry corresponding to the non-supersymmetric 7-branes that we found in our paper turns out to be written in terms of the functions that we introduced in the previous section as
\begin{equation}
ds_{12}^2 \,=\, \left(\frac{\sinh \eta}{D} \right)^\frac{1}{4}ds^2_{\text{Mkw}_{8}}+\frac{e^{h_0(\eta,\Theta)}}{64}\frac{1}{D^\frac{9}{8}(\sinh \eta)^\frac{7}{8}}(d\eta^2+d\Theta^2)
  \,+\, \frac{\left|dy-\tau dx \right|^2}{\mathrm{Im}(\tau)} \ ,
\end{equation}
with
$D=\cosh \eta-\cos \Theta$, $h_0$ given by \eqref{h0final},
and $\tau=A( i \sech \lambda - \tanh \lambda)$, in terms of the specific $\lambda$ given in \eqref{lambdafinal} (or $h_0$ given by \ref{secondh0} and $\lambda$ given by \ref{secondlambda}).

This 12D geometry is an extremum of the pure Einstein-Hilbert action and as such, it is Ricci-flat. Geometries of this type should be interpreted as non-supersymmetric gravitational monopole sources, just in the same way as the D7-brane background lifts to a 12D KK-monopole, \emph{i.e.} a Euclidean Taub-NUT (ETN) times flat space. The exact nature of these objects in our case still remains a bit obscure, but it is certainly beyond the known KK-monopole geometry, due to the non-holomorphicity of $\tau$. Bearing in mind the analogy with the 12D description of D7 branes, which turns out to be $\mathrm{Mkw}_8\times \mathrm{ETN}_4$, we expect the singularities  which are generically present on the unit disk in the $(\eta,\Theta)$ plane to be resolved into a smooth 12D spacetime.

Finally, one might say that our construction of general 7-brane backgrounds may be used as an explicit recipe to engineer novel non-supersymmetric Ricci-flat geometries that do not rely on a holomorphic axio-dilaton. However, it is important to stress in this context that global issues may arise depending on the specific 7-brane monodromy that we choose. In particular, when choosing the R7 brane monodromy, the $\mathbb{T}^2$ topology gets modified into a Klein bottle. This fact is due to the loss of orientability following from reflection monodromy along $S^1_{\Theta}$. For a safe global interpretation, one should be careful about issues of orientation and, for example, (s)pin structures. A complete and exhaustive discussion remains beyond the scope of our present work.

\section{Conclusion \& Outlook}\label{conclusions}
In this work, we have explored the connection between the AdS$_9$ solutions of type IIB supergravity and R7-branes. We rewrote the AdS$_9$ backgrounds of \cite{Dibitetto:2026oor} in a particularly simple form and used this parametrization to construct a family of exact 7-brane solutions whose near-horizon limit reproduces them. We then analyzed the global properties of these solutions and showed that imposing the reflection monodromy of the axio-dilaton selects a distinguished subclass, which we proposed to interpret as the gravitational description of R7-branes. The same parametrization also allowed us to derive an exact expression for the holographic central charge and to formulate a 12D geometric description of the 7-brane solutions in terms of Ricci-flat solutions.

The important open question is whether the AdS$_9$ solutions discussed in this work admit an 8D holographic dual CFT. The results presented here provide a first step towards answering this question by proposing a concrete 7-brane realization of these backgrounds. Nevertheless, it remains unclear whether these solutions indeed admit a holographic interpretation and, if so, what the corresponding microscopic quantum degrees of freedom are. From this perspective, the interpretation of the holographic central charge \eqref{chol} is particularly intriguing. At the moment, its microscopic origin remains completely unclear. The puzzle becomes even more striking in the limit described by \eqref{AdS9_C0_SOL}, where the axion vanishes. Understanding the brane interpretation of the solution \eqref{AdS9_C0_SOL} therefore appears to be an essential step towards identifying the quantum degrees of freedom underlying these backgrounds.

An interesting direction for future work would be to explore the possibility that our non-supersymmetric 7-brane solutions should be understood as bound states of 7-branes and D-instantons. Such bound states would represent genuinely new non-perturbative objects, whose physical properties cannot be reduced to those of their individual constituents. This picture could provide the right quantum degrees of freedom needed to explain the holographic central charge and, more generally, help clarify the existence of an eight-dimensional holographic dual for the AdS$_9$ solutions.

Another intriguing trajectory to explore is whether there is a more general class of exact 7-brane solutions, perhaps with different near-horizon geometries, bringing us closer to a classification of co-dimension two defects in supergravity. The Ernst equation \eqref{ernsteq} has a wide variety of interesting solutions in the literature (see, for example, \cite{ernstsols}) which we expect to lead to a rich variety of exact 7-brane solutions with the appropriate monodromy.

Finally, we are well aware that the absence of supersymmetry raises important questions about the stability of our solutions. While this issue lies beyond the scope of the present work, understanding their perturbative and non-perturbative stability will be essential. We hope to return to these questions in future work.
\\
\\
\\

\section*{Acknowledgements}

We thank Atakan \c{C}avu\c{s}o\u{g}lu, Mirjam Cveti{\v c}, Jonathan Heckman, Jeffrey Kuntz and Chitraang Murdia for bringing to our attention a redundancy between the R7-brane solution presented in the first version of this paper and the AdS$_9$ near-horizon geometry. This issue has been resolved in the present version, where we construct genuinely new R7-brane solutions.
\newline

We would like to thank M. Akhond, M. Bianchi, Atakan \c{C}avu\c{s}o\u{g}lu, Mirjam Cveti{\v c}, Okan G\"unel, Daniel Junghans, Jacob McNamara, Ethan Torres, A. Sagnotti, R. Savelli, and Cumrun Vafa for interesting discussions and comments. The work of GD is partly supported by an INFN fellowship under the “Iniziativa Specifica” ST\&FI. C.S. is supported by a Postdoctoral Research Fellowship granted by the F.R.S.-FNRS (Belgium).

 \bibliographystyle{utphys}
  \bibliography{references}
\end{document}